# The Hall Effect and Hole Densities in High $T_C$ $Ga_{1-x}Mn_xAs$ Thin Films

K.W. Edmonds, K.Y. Wang, R.P. Campion, A.C. Neumann, C.T. Foxon, B.L. Gallagher, and P.C. Main

*School of Physics and Astronomy, University of Nottingham, Nottingham NG7 2RD, UK*

By studying the Hall effect in a series of low resistivity $Ga_{1-x}Mn_xAs$ samples, accurate values for the hole density $p$, Mn concentration $x$, and Curie temperature $T_C$ are obtained over the range $0.015 \leq x \leq 0.08$. The hole density corresponds to 90% of the Mn concentration at low $x$, and has a maximum value of $1.0 \times 10^{27} m^{-3}$ when $T_C=125K$ for $x=0.06$. This data allows the first meaningful comparison of mean field predicted Curie temperatures with experiment over a wide range of $x$. The theory is in qualitative agreement with experiment, but overestimates $T_C$ at large $x$ and underestimates $T_C$ at low $x$.

Ferromagnetic semiconductors have enormous potential for 'spintronics' device applications [1]. Ferromagnetic $Ga_{1-x}Mn_xAs$, in which the exchange interaction is mediated by mobile holes, is the most widely studied system both experimentally and theoretically [2,3]. Progress towards understanding this material and, in particular, the mechanisms limiting the Curie temperature $T_C$ (the current limit is $\approx 110K$ [1]), is dependent on a knowledge of the hole density, $p$. However, the dominant anomalous Hall effect means that accurate determination of $p$ is only possible if the magnetoresistance is weak, i.e. in samples showing metallic conduction at low temperatures [4]. Previously, this has been limited to a narrow region of Mn concentrations, $x$, close to 0.05 [1]. By limiting the film thickness to 45nm, and carefully controlling growth and post-growth annealing conditions, we have extended the range over which samples show metallic conduction to $0.015 \leq x \leq 0.08$, and obtained Curie temperatures exceeding 110K. This allows us to obtain hole densities over a much wider range of $x$ and $T_C$ than has been reported previously. We show that, in contrast to expectations, the hole density is not necessarily a small fraction of the Mn concentration. These measurements provide a test ground for comparison of transport and magnetic properties with the predictions of the existing theories of ferromagnetism in GaMnAs.

The Hall resistivity in magnetic materials is given by

$$\rho_{xy} = \rho_{xy}^o + \rho_{xy}^a = R_0 B + R_a(\rho_{xx}) M \qquad (1)$$

where the normal contribution, $\rho_{xy}^o$, is proportional to the external magnetic field $B$, $R_0=1/pe$, and the anomalous contribution, $\rho_{xy}^a$, is proportional to the macroscopic magnetisation $M$ [5]. $R_a$ arises from the spin-orbit interaction, which induces anisotropy between scattering of spin-up and spin-down electrons. Particular 'skew' and 'side-jump' scattering mechanisms predict $R_a \propto \rho_{xx}$ and $R_a \propto \rho_{xx}^2$ respectively, where $\rho_{xx}$ is the longitudinal resistivity [5]. Recently, it has been suggested [3,6] that the dominant contribution to $R_a$ in Ga$_{1-x}$Mn$_x$As is due to a scattering-independent topological contribution, which also gives $R_a \propto \rho_{xx}^2$. To determine $p$ from $\rho_{xy}$, the two contributions must be separated reliably. This is complicated by the magnetoresistance of $\rho_{xx}$ at low temperatures, and the large paramagnetic response of $M$ at temperatures above $T_C$. An accurate determination of $p$ from Hall effect measurements has previously only been possible at $x=0.053$.

Our Ga$_{1-x}$Mn$_x$As films, with 1.5% to 8% Mn, were grown on semi-insulating GaAs(001) substrates by low temperature (180ºC-300ºC) molecular beam epitaxy using As$_2$. The samples were grown at the highest temperatures possible while maintaining two-dimensional growth [7]. Two Ga sources were used in order to maintain stoichiometry and growth temperature during growth. The film thickness, $d$, was chosen to be 45nm in order to minimise the effect of the variation of the growth parameters during deposition and maximise $T_C$ [8]. Mn concentrations were determined from the Mn/Ga flux ratio, and confirmed by x-ray fluorescence measurements. Full details of the growth and structural characterisation are presented elsewhere [7].

Hall bar structures were fabricated lithographically from the Ga$_{1-x}$Mn$_x$As thin films. Magnetoresistance and Hall resistance measurements were performed simultaneously using standard low frequency ac techniques, in magnetic fields between ±16T and temperatures down to 0.3K.

The temperature dependence of the resistivity for the Ga$_{1-x}$Mn$_x$As samples is shown in figure 1. The peaks in these curves correspond approximately to the Curie temperature $T_C$. The precise value of $T_C$ is obtained from Hall measurements, as discussed below. Low resistivity 'metallic' samples, with a positive (d$\rho_{xx}$/dT) below $T_C$, were obtained for $0.015 \leq x \leq 0.08$. This is in contrast to previous studies where samples with $x \leq 0.03$ or $x \geq 0.06$ were insulating at low temperatures [1]. It has been suggested that this is a fundamental limit to the metallic phase [9]; however, the present result shows that this is not the case. Annealing of Ga$_{1-x}$Mn$_x$As close to the growth temperature is known to give an improvement in sample quality [11-13]. Applying this procedure to our samples resulted in an increase of $T_C$ to 125K and 112K respectively for $x=0.06$ and 0.08, comparable to or larger than the highest reported values [1,13], as well as a decrease in the low temperature resistivity. Resistivities at 4.2K for both as-grown and annealed samples are shown in figure 3a.

Since $\rho_{xy}^a \gg \rho_{xy}^o$, and $\rho_{xx}$ is usually weakly dependent on $B$, the functional form of $\rho_{xy}(B)$ should follow that of $M(B)$ at low fields. The inset of figure 2 shows a typical hysteresis curve of $\rho_{xy}$ vs. $B$. The temperature dependence of the saturation magnetisation $M_S$ can be obtained from measurements of $\rho_{xx}$ and $\rho_{xy}$ using equation

(1) and the Arrott plot method [1]. Figure 2 shows $M_S(T)$ for the annealed $Ga_{0.94}Mn_{0.06}As$ sample with $T_C=125K$, obtained by assuming either a linear or quadratic dependence of $\rho_{xy}$ on $\rho_{xx}$. The measured form of $M_S(T)$ depends on the assumption made about the dependence of $R_a$ on $\rho_{xx}$; however, accurate values of $T_C$ can be obtained by this method since $M_S$ is varying rapidly close to $T_C$ while $\rho_{xx}$ is varying relatively slowly. Values of $T_C$ obtained in this way are shown in figure 3b.

To obtain hole densities from Hall measurements one must carry out measurements in the limit where the magnetisation is fully saturated i.e. at low temperatures and high magnetic fields [4]. However, even in this limit $\rho_{xy}^a$ is not constant due to the dependence of the coefficient $R_a$ on the longitudinal magnetoresistance. The behaviour of $\rho_{xx}$ and $\rho_{xy}$ out to high magnetic fields is shown in figure 4 for a $Ga_{1-x}Mn_xAs$ sample with x=0.05. For all samples measured, $\rho_{xx}$ shows an initial positive magnetoresistance (MR), as is typically observed for $Ga_{1-x}Mn_xAs$ with in-plane magnetic anisotropy [1], followed by a negative MR beyond B≈0.5T. The magnitude of the observed negative MR correlates with $\rho_{xx}$, and is found to be largest in samples that are insulating at low temperatures. In the present samples the negative MR is typically of order 3% at 10T and 4.2K, weaker than has been reported previously [1]. A small MR is crucial for obtaining accurate values for the hole density in this system.

$\rho_{xy}$ shows a rapid rise at low B due to magnetic saturation of the ferromagnetic film, followed by a more gradual rise that has contributions from both $R_o$ and $R_a$. Fitting the measured $\rho_{xy}$ versus B curve using eqn. (1) yields values for $p$ which differ by 15% for this sample depending on whether $R_A \alpha \rho$ or $R_A \alpha \rho^2$. The fitted curves are shown in figure 4.

$\rho_{xy}$ and $\rho_{xx}$ were measured at high magnetic fields (10-16T) as a function of temperature over the range 0.3-7.0K. This allows the determination of the functional form of $R_A(\rho_{xx})$. We find $R_A \alpha \rho_{xx}^n$, where $n$ is sample dependent, but for all samples $1<n<2$. This relationship can then be used to fit the $\rho_{xy}$ versus B curves and extract accurate values for the hole density. Further details of this fitting procedure are presented elsewhere [14].

The hole densities determined from this fitting procedure are shown per unit volume in figure 3c, and per Mn ion in figure 3d. In the as-grown samples, $p$ shows a broad peak at $x=0.05$, and is sharply decreased for $x=0.08$. The low temperature annealing of $x=0.06$ and $x=0.08$ samples results in an increase of $p$ by a factor of 2.3 and 5.0 respectively. Such behaviour has been ascribed to a decrease in the concentration of Mn donors at interstitial sites [15]. The number of holes contributed per Mn falls approximately linearly with $x$, suggesting an increasing concentration of compensating defects. For low doping ($x<0.03$) we observe an unexpectedly large hole concentration, equal to around 90% of the Mn concentration. Magnetism in such weakly compensated samples has been given little consideration theoretically.

From the measured low temperature resistivities and hole densities, the mean free path $\Lambda$ of holes in $Ga_{1-x}Mn_xAs$ can be determined. This is shown in figure 3e. We obtain $\Lambda$=6-8Å for $x\leq0.06$, falling to lower values for $x=0.08$. These values are slightly smaller than the average Mn ion separation (~1.3-0.9nm for $x$=0.02-0.06). Such low values of $\Lambda$ are expected for the low temperature grown $Ga_{1-x}Mn_xAs$ due to the high degree of disorder in this system. Annealing leads to a marked increase of $\Lambda$, which is a consequence of the reduction of disorder. This is also likely to contribute to the increase in Curie temperature to over 110K for the annealed samples.

Within the mean field model discussed in refs. [2,3,9], the Curie temperature in dilute magnetic semiconductors is determined by the values of $p$ and $x$. Therefore, the present data can be used as a test of this widely used theory, which has previously only been applied for the value $x$=0.053. Mean field predicted values of $T_C$ obtained from the known values of $p$ and $x$ are shown by squares in figure 3b. Additional parameters are as described in ref. [2]. It can be seen that the mean field model overestimates $T_C$ for $x \geq 0.04$. However, for $x$=0.02, the measured $T_C$ is substantially *larger* than the mean field prediction. This is surprising, since the mean field description neglects disorder and the effect of long range fluctuations, and so could be considered as an upper limit to $T_C$ [16]. This may be evidence that the presence of disorder can lead to an increase of $T_C$, as predicted theoretically [10]. The mean field theory implicitly assumes that the Fermi wavelength is much larger than the separation of Mn ions, and thus is most applicable in the case where compensation is large. However, this condition may be met even when the hole / Mn ratio is close to 100% because of the degeneracy of the heavy hole and light hole bands [3].

In summary, the Hall effect has been carefully studied in a series of metallic $Ga_{1-x}Mn_xAs$ samples with $0.015 \leq x \leq 0.08$. This gives accurate values for the hole density, Curie temperature, and Mn concentration over a much wider range than has previously been possible. We observe that, contrary to expectations, the hole density can be as much as 90% of the Mn concentration. In addition, we have conducted the first meaningful comparison of mean field predicted Curie temperatures with experiment over a wide range of $x$.

**Acknowledgements**

The project was funded by EPSRC (GR/R17652/01) and the European Union (FENIKS project EC:G5RD-CT-2001-00535). We thank Jaz Chauhan and Dave Taylor for processing the Hall bars.

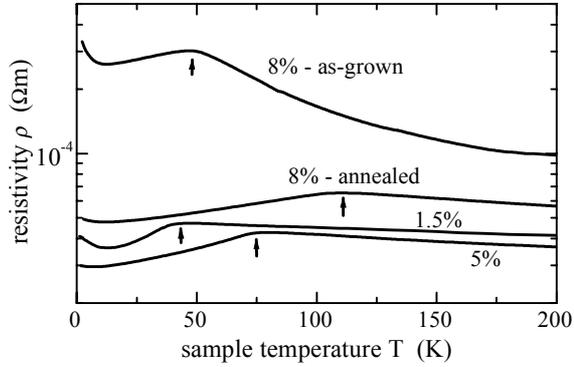

FIG. 1. Resistivity versus temperature for $Ga_{1-x}Mn_xAs$ films with x=0.015, 0.05, 0.08 as-grown, and x=0.08 after annealing. The arrows mark the $T_C$ values for each sample.

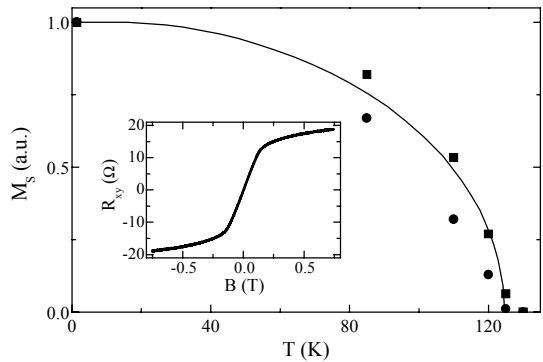

FIG. 2. Saturation magnetisation versus temperature assuming $R_a \alpha \rho_{xx}$ (circles) and $R_a \alpha \rho_{xx}^2$ (squares), obtained from the anomalous Hall effect for an annealed $Ga_{0.94}Mn_{0.06}As$ sample with $T_C$=125K. Inset: hysteresis curve of Hall resistance versus perpendicular external magnetic field for this sample at T=120K.

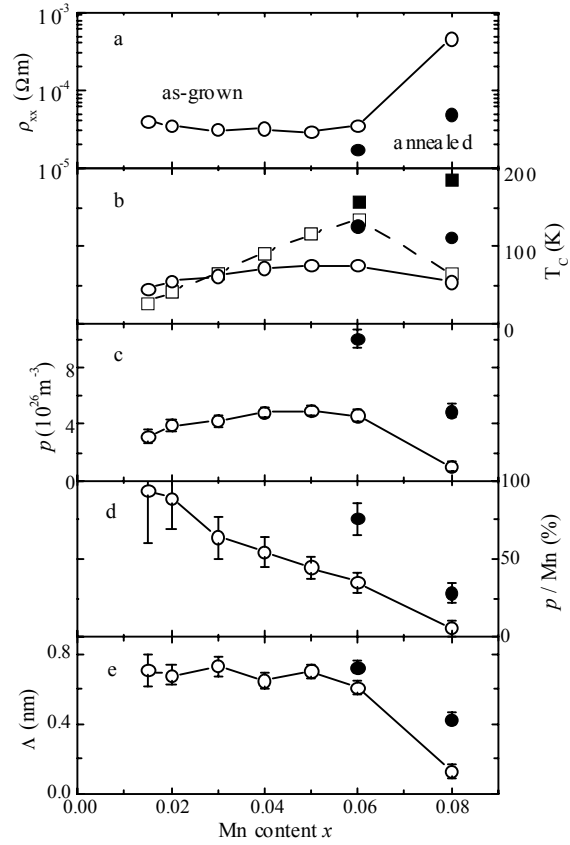

FIG. 3. $Ga_{1-x}Mn_xAs$ sample parameters determined from transport measurements versus Mn content x: (a) resistivity at 4K; (b) Curie temperature $T_C$; (c) hole density per cubic metre; (d) hole density per Mn; (e) mean free path. Open circles are as grown, full circles are after annealing. Mean field predicted $T_C$ are shown in figure 3b for as-grown (open squares) and annealed (full squares) samples.

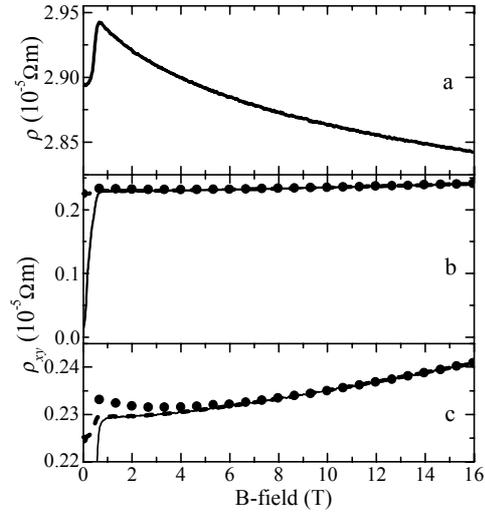

FIG. 4. (a) resistivity $\rho_{xx}$ and (b) Hall resistivity $\rho_{xy}$ versus external field for $Ga_{0.95}Mn_{0.05}As$. $\rho_{xy}$ Fitted curves with $R_a \propto \rho_{xx}$ and $R_a \propto \rho_{xx}^2$ are shown by dashes and circles respectively. Data in (b) is shown on a magnified scale in (c).